# On the applications of μ=-1 metamaterial lenses for magnetic resonance imaging

Manuel J. Freire*, Lukas Jelinek, Ricardo Marques
Departamento de Electronica y Electromagnetismo. Universidad de Sevilla.
Facultad de Fisica, Avda. Reina Mercedes s/n. 41012 Sevilla, SPAIN

*Corresponding author.
Fax: +34954239434
Email address: freire@us.es
URL: http://alojamientos.us.es/gmicronda/Miembros/Freire/freire.htm

## Abstract

In this work some possible applications of negative permeability magnetic metamaterial lenses for magnetic resonance imaging (MRI) are analyzed. Metamaterials are artificial composites designed to have a given permittivity and/or permeability, including negative values for these constants. It is shown that using magnetic metamaterials lenses it is possible to manipulate the spatial distribution of the radio-frequency (RF) field used in MR systems and, under some circumstances, improve the sensitivity of surface coils. Furthermore a collimation of the RF field, phenomenon that may find application in parallel imaging, is presented. MR images of real tissues are shown in order to prove the suitability of the theoretical analysis for practical applications.

## 1. Introduction

Metamaterials are artificial composites whose electromagnetic properties can be engineered to achieve extraordinary phenomena not observed in naturalmaterials as, for instance, negative effective permittivity and/or permeability [1]. Effective permittivity and permeability of metamaterials arise from their structure rather than from the nature of their components, which usually are conventional conductors and dielectrics. Metamaterials are fabricated by means of the repetition of a resonant element to constitute a periodic structure. An essential characteristic of metamaterials is that both the size of this element and the periodicity are smaller than the wavelength of the electromagnetic fields that propagate through the structure, so that an effective permittivity and permeability can be defined through the appropriate homogenization procedure [2]. One of the most striking properties of metamaterials is the ability of a metamaterial slab with relative permittivity ε and relative permeability μ, both equal to -1, to behave as a "super-lens" with sub-wavelength resolution [3], that is, with a resolution smaller than the free-space wavelength of the impinging radiation. Although this effect is severely limited by losses, it is now well supported by many experiments and theoretical calculations (see, for instance, [2] and references therein). Interestingly,
if the frequency of operation is sufficiently low, as it happens in MRI, we are in the realm of the quasi-statics, and we only need a metamaterial slab with ε = −1 or μ = −1 (depending on the electric or magnetic nature of the quasi-static field) in order to observe this effect [3]. Therefore, if we place a μ = −1 metamaterial slab between a RF magnetic field source (for instance an oscillating magnetic dipole) and a receiving device (for instance a MR surface coil) the slab will image the source towards the receiver, which will "see" the source closer than it actually is, thus detecting a stronger signal. It is clear that this mechanism can be applied to improve the sensitivity of MR surface coils as far as the additional noise introduced by the device will not compensate the gain in the signal. As it has been just explained, µ = −1 metamaterial slabs have the ability to virtually "approach" the source to the detector. As it will be shown in this paper, this can be useful not only to improve the signal, but also to provide a better localization of the field of view (FOV) of the detector, with potential applications in parallel imaging. Of course, in order to take advantage of all these capabilities, metamaterials should not interact with the static magnetic fields used to codify the oscillating magnetic dipoles in MRI. Fortunately, as it was already mentioned, metamaterials are usually made of conventional dielectric and conductors, so that the compatibility with static MR magnetic fields can be achieved by using nonmagnetic conductors. The interesting properties of most metamaterials occur in a very narrow band of frequencies due to the resonant nature of the elements that constitute the periodic structure. This narrow bandwidth is usually cited as one of the main limitations for metamaterial applications. However, it is not a problem for MRI applications, because MR images are acquired by measuring RF signals inside a relatively narrow bandwidth of a few tens of kilohertz. In addition, since the wavelength associated with RF fields is of the order of the meters, it is possible to use conventional printed circuit techniques to develop quasi-continuous metamaterials with constituent elements and periodicities two orders of magnitude smaller than the wavelength. Application of metamaterials in MRI has been explored previously in several works [4]-[11]. Basically, two types of metamaterials which correspond to two different resonant elements have been used for MRI applications. The first group are swiss roll metamaterials [4]-[8]. A swiss roll consists of a conductive layer which is wound on a spiral path around a cylinder with an insulator separating consecutive turns. Reported experiments [4, 5] proved that swiss-roll metamaterials can guide the RF flux from a sample to a remote coil. The guiding behavior is due to the high effective permeability of the metamaterial. The reported experiments also proved that these metamaterial guides can be employed in imaging [4] and spectroscopic [5] experiments for excitation as well as reception. In all these applications, the swiss rolls metamaterials mimic a medium with very high magnetic permeability at the proper frequency. A 2D log-pile structure of swiss rolls, which mimics a two-dimensional µ = −1 medium was also used to demonstrate sub-wavelength imaging of a pair of parallel wire currents [6], but no direct application to MRI of this device has been reported up to the date. The other group of metamaterials which have been applied to MRI are capacitively loaded split ring metamaterials [9]-[11]. A capacitively loaded split ring is a small open ring of copper which is loaded in the gap with a capacitor. Of course, this capacitor has to be non-magnetic for MRI applications. Split rings have the key advantage over swiss rolls of providing three-dimensional (3D) isotropy when they form a cubic lattice [9, 10], which is an essential property if the device has to image 3D sources. A split ring is similar to a very small parasitic MR coil. However, whereas a MR coil works at resonance, the working frequency of split rings in a µ = −1 metamaterial lens differs from its frequency of resonance [9, 10], which also helps to reduce losses and noise. It is the collective behaviour of split rings at this working frequency, which is different to the resonant frequency, what provides the relative effective permeability equal to -1 [10]. Split rings were used as the constituent elements of a 3D lens that was fabricated and tested in a 1.5T MRI system [9]. Fig. 1 shows a photograph of this device. Almost simultaneously, an accurate model for this design was developed, which showed the consistency of the continuous medium description of the device [10]. The goal of this design was to provide a permeability μ = −1 at the Larmor frequency of the MRI system. The reported experiment [9] showed the capability of a μ = −1 metamaterial lens to improve the signal detected by a surface coil for a particular configuration, but did not provide a systematic analysis of the capabilities of such metamaterial lenses for MRI applications. The main aim of this work is to provide a first approximation to this analysis. The paper is organized as follows: first, a theoretical model based on a continuous medium approach is developed. Using this model, the sensitivity, the signal-to-noise ratio (SNR) and the FOV localization provided by μ = −1 metamaterial lenses are analyzed in several situations. Inspired on this analysis some experiments were designed and carried out in order to test the conclusions derived from the analysis. Finally, potential future applications of metamaterial lenses for MRI will be discussed.

## 2. Model

The main aim of this section is to develop a method for the computation of the signal, the noise, and then the SNR detected by a surface coil in the presence of a metamaterial slab placed at several distances. The analysis will include configurations that mimic the experiments reported in [9] and [11], as well as other potentially useful configurations. Through the analysis it will be assumed that the metamaterial slab behaves as an effective homogeneous medium with some effective permittivity and permeability. To demonstrate that the device shown in Fig. 1 actually behaves as an effective homogeneous medium is outside the scope of this paper, since this was shown in detail in our previous work [9]-[11]. Although the analysis could be done, in principle, by using a conventional electromagnetic solver, our experience is that these solvers have difficulties to deal with effective media of negative permeability, particularly those close to μ = −1. Therefore, a specific code has been developed for this analysis. In order to simplify the following discussion, Fig. 2 shows the structure under analysis. This structure includes a coil of average $\rho_0$ made of a lossless metallic strip of width $w$ and negligible thickness. The coil is placed at certain distance of a piecewise homogeneous multilayered medium with the coil axis perpendicular to the layers. The layers have a thickness $t$ and an arbitrary complex permittivity ε and permeability μ, and depending on the values of these parameters they can model either the μ = −1 lens, a specific tissue, or air. This simplified model has the advantage of an easy analytical solution, and we feel that it retains the most salient features of real experiments, at least qualitatively.

The first step in the analysis is the computation of the signal received by the coil. According to reciprocity theorem this signal is proportional to the $B_1$ magnetic field produced by the detector when it is driven by a unit current [12]. We begin with the calculation of the B1 field by considering a current density given by

$$J_\varphi(\rho, z) = K_\varphi(\rho)\delta(z) \tag{1}$$

where $K_\varphi(\rho)$ is the surface current density on the ring. This surface current density is approximated by a Maxwellian distribution in the ring cross-section as

$$K_\varphi = K_0 \Big/ \sqrt{1 - \left(\frac{\rho - \rho_0}{w/2}\right)^2} \tag{2}$$

which gives quite good approximation for the actual current distribution on the metallic strip provided that the ring radius is not too small. In order to compute the magnetic field, the equation for the vector potential **A** must be solved

$$\Delta A + k^2 \mathbf{A} = -\mu \mathbf{J} \tag{3}$$

where k accounts for the wavenumber. This equation has to be solved at each layer and the boundary conditions have to be satisfied. Since the studied system has angular symmetry, Eq.(3) can be easily solved with the help of a Hankel transformation of the first order. Particularly, in the layer containing the coil, the transformed equation has the following shape

$$-\frac{\partial^2 \tilde{A}_\varphi}{\partial z^2} - k_z^2 \tilde{A}_\varphi = \delta(z) \mu \tilde{K}_\varphi \tag{4}$$

where

$$k_z^2 = k^2 - k_\rho^2 \tag{5}$$

$k_\rho$ is the spectral wavenumber associated with the transverse coordinates x, y, and the tilde above the quantities denotes Hankel transformation. Eq.(4) can be easily solved giving

$$\tilde{A}_\varphi(k_\rho, z) = -\frac{j\mu}{2k_z} e^{-jk_z|z|} \tilde{K}_\varphi(k_\rho) + C^-(k_\rho) e^{jk_z z} \tag{6}$$

where the first summand in the second term of Eq.(6) represents the excitation, and the second one represents the reflected wave. In the remaining layers, the corresponding equation is the expression in Eq.(4) without the source term, and the solution is

$$\tilde{A}_\varphi = \left( C^+(k_\rho) e^{-jk_z z} + C^-(k_\rho) e^{jk_z z} \right) \tag{7}$$

The unknown coefficients $C^+(k_\rho)$, $C^-(k_\rho)$ can be determined for each layer by imposing the appropriate boundary conditions, specifically, by enforcing the continuity of the tangential components of the electric and magnetic fields on each boundary. After the unknown coefficients $C^+(k_\rho)$, $C^-(k_\rho)$ have been determined, the magnetic field is know at every point in space. As it was already mentioned, the signal is proportional to the $B_1$ magnetic field produced by the coil when it is driven by a unit current. On the other hand, the MR noise is proportional to the square root of the noise resistance R associated with the sample [13]. In our analysis the coil is assumed to be lossless, which means that both the coil losses and the MRI system losses are excluded from the analysis. Therefore the computed noise will be a sort of intrinsic noise [14]. Since we are interested in the comparison of the SNR given by different configurations, taking into account that the SNR is proportional to $B_1/\sqrt{R}$, in our analysis we will compute and compare this quantity for the different configurations. This quantity can be seen as a sort of normalized SNR. For the sake of simplicity, we will term SNR to the quantity $B_1/\sqrt{R}$, but the previous considerations must be taken into account in order to get a correct interpretation of the results. Once the magnetic field is calculated, the next step is the computation of the noise resistance *R*. Usually, this resistance is calculated from the power dissipated by the eddy currents Js induced in the sample as:

$$R = \frac{1}{|I|^2} \text{Re}\left[\int_V \mathbf{J}^s \cdot \mathbf{E}^c \mathbf{dv}\right] \quad (8)$$

where $\mathbf{E}^c$ is the electric field induced by the coil, and *I* is the coil intensity, which is set equal to unity. In our analysis, we use reciprocity theorem ([15], pp.116) to obtain this resistance from the reaction between the current in the coil Jc and the electric field reflected by the sample $\mathbf{E}^r$, which is defined as the field created by all currents in the multilayer medium (that is, all currents in the system, except the imposed current on the coil):

$$R = -\frac{1}{|I|^2} \text{Re}\left[\int_V \mathbf{J}^c \cdot \mathbf{E}^r \mathbf{dv}\right] \quad (9)$$

Particularly, for the case presented above, this resistance can be calculated as

$$R = -\frac{2\pi}{|I|^2} \text{Re}\left[\int_0^\infty K_\varphi \cdot E_\varphi^r \, \rho \, d\rho\right] \quad (10)$$

which can be advantageously rewritten using Plancherel-Parseval theorem as

$$R = -\frac{2\pi}{|I|^2} \text{Re}\left[\int_0^\infty \widetilde{K}_\varphi \cdot \widetilde{E}_\varphi^r \, k_\rho \, dk_\rho\right] \quad (11)$$

Once the noise resistance *R* has been computed, the normalized SNR given by $B_1/\sqrt{R}$, can be readily computed, which ends our analysis.

## 3. Results

### 3.1. Numerical results

Following the method described above, the signal and the SNR for a surface coil of 3 inch in diameter and a strip width of 1 cm, have been computed at the frequency of 63.87 MHz corresponding to the Larmor frequency of a 1.5 T MRI system, which is the type of system used in the experiments reported in this work. According to the previous analysis, the sample is modelled as a lossy layer with a complex permittivity whose real and imaginary parts are given by the mean values corresponding to human tissues [16]. The lens is modelled as a layer with a complex permeability whose imaginary part accounts for the losses. Fig. 3 shows on the left side a sketch of the two different configurations analyzed in this Section. In one of them the coil is placed at certain distance d of a semi-infinite lossy media (no lens case in the figure) which models the sample. In the other configuration the coil is placed at the same distance d from the lens and this lens is in contact with the sample. In our calculations the lens was modelled as a 3 cm thick slab with a complex permeability $\mu = -1 - j0.25$, which corresponds to the realistic values calculated for the lens shown in Fig. 1 from previous developed models [10, 17]. The sample was modelled by using an average value for the permittivity of human tissues, $\varepsilon = 90 - j197$, whose imaginary part corresponds to a conductivity of 0.7 S/m at 63.87 MHz [16]. Fig. 3a shows the signal (the axial magnetic field) along the coil axis for a distance $d = 1$ cm. The solid line in Fig. 3a (no lens case) corresponds to the signal provided by the coil in the absence of the lens. The dashed line (realistic lens case) corresponds to the signal provided by the coil with the lens placed between the coil and the lossy media. Finally, the dotted line (lowloss case) corresponds to the signal provided by the coil in front of a hypothetical lens with an imaginary part of the permeability one order of magnitude smaller than for the realistic lens. As expected, the comparison between the different curves shows that the signal with the lens is always larger that the signal without the lens for the same distance $d$. The dip in the signal observed for the lowloss case is due to the strong oscillations of the magnetic field at the output interface of the lens [2]. Fig. 3b shows the signal for a distance $d = 6$ cm, with a qualitative behavior similar to that shown in Fig. 3a. The normalized SNR corresponding to the cases analyzed in Fig. 3a and 3b are shown in Fig. 3c and 3d, respectively. It can be seen that for the smaller distance, d = 1 cm, the SNR is worse (but still similar) in the presence of the lens than in the absence of it. However, for larger distances, such as $d = 6$ cm, the SNR becomes better when the lens is present (numerical computations not shown here prove that this improvement is even better for higher distances). It is worth to mention that this behavior is almost unchanged when losses in the lens are reduced by an order of magnitude, as it is shown in Fig. 3d. Therefore, reducing lens losses - for instance, using superconducting split rings [18] - is not enough to improve the SNR over the no lens case when the coil is near the lens (and the lens near the sample). From field computations we have realized that this fact is related to the very high values of the induced fields at the interface between the lens and the sample that appear in this configuration. These strong fields cause a strong dissipation in the sample, and therefore a substantial increase of noise. The presence of such strong fields at the lens-sample interface is a well known effect in low-loss metamaterial lenses (see, for instance, [2] and references therein) which is related to the strong variations of the magnetic field at the input lens interface (the interface closer to the coil) when the coil is near this interface. From the results reported in Fig. 3 and from the above considerations, it can be concluded that the presence of the lens always improves the signal, but only improves the SNR when the coil is placed at some distance of the lens. When the coil is placed near the lens, the SNR is not improved by the lens, although it remains of the same order as in the absence of the lens. From systematic numerical calculations and experiments (not shown in this work) it is concluded that, as a rule of thumb, the distance between the coil and the lens should be at least equal to the diameter of the coil in order to ensure an improvement of the SNR. Since the configuration in the experiment previously reported by the authors [9] fulfils this requirement, the analysis is now oriented to a configuration that resembles this experiment. The experiment consisted of placing the lens between the knees of a volunteer in order to image both knees with a 3-inch coil placed near one of the knees. Fig. 4a shows in logarithmic scale the normalized SNR for the same coil geometry as in Fig. 3 but with the lens placed between two lossy slabs (noted as tissue in the figure) which may represent the knees in the reported experiment [9] or any other tissue. The curves in the figure show that the SNR in the slab closer to the coil is the same regardless the lens is present or not. However, the SNR is highly improved for the slab beyond the lens, as it is shown in detail in Figs. 4b and 4c. These figures show a 2D plot of the SNR in a plane perpendicular to the coil and containing its axis when the lens is present - Fig. 4b - and when the lens has been replaced by an air layer of the same thickness - Fig. 4c. The reported results clearly show the improvement of the SNR when the lens is placed between the lossy layers. Since the sensitivity of surface coils strongly decays with distance, it will be useful if, on the basis of the previous concepts, it can be developed some way of reducing this decay. This can be possible by backing the lens with a magnetic wall, that is, a medium with a large value of the permeability (which, in practice, could also be another metamaterial slab). According to image theory in electromagnetics, this configuration will create an "image" beyond the magnetic wall composed by an identical lens and an identical coil, carrying the same current as the original one. Therefore, the magnetic field (and the signal) in the region of interest will be increased by the presence of these additional coil and lens. Fig. 5a shows the SNR along the coil axis for this configuration. Figs. 5b and 5c show 2D plots of the SNR in a plane containing the coil axis. In Fig. 5b the lens is backed by a magnetic wall. In Fig. 5c both the lens and the magnetic wall have been removed. The comparison between both figures makes clear the increasing of the SNR in the region between the coil and the lens at distances where the coil sensitivity has decayed appreciably in the absence of the lens. This result is experimentally checked in the following section. Probably the most studied property of metamaterial lenses is their ability to improve the discrimination between the fields coming from two independent sources [3]. Translated to MRI terminology, this property implies an improved localization of the FOV of each coil in an array of surface coils, a fact that could find application in parallel imaging [19]. Some numerical computations carried out using the method described in the previous section are shown in Fig. 6. This figure shows the B1 field produced at 6 cm inside a lossy semi-infinite media by two rectangular coils with dimensions of $7 \times 23$ cm and with their centers separated 10 cm. These dimensions correspond to the coil geometry used in [19]. The coil array is placed at 1.5 cm from the sample and at the same distance of the lens analyzed in this work, which is then placed on the sample, as it is shown in Fig. 6a. The results in Fig. 6b and 6c show that the field pattern of each coil can be distinguished much better when the lens is present than when it is not. In other words, the lens improves the localization of the FOV of each coil in the array. As it is expected from the previous results - Fig. 3c - this should not imply a significant loss in the SNR.

## 3.2. Experimental results

First, for the purpose of illustrating how a $\mu = -1$ metamaterial lens does really transfer beyond the lens the field pattern of a source, Fig. 7 shows a sketch of a simple experiment and the MR images obtained in this experiment where a coil of 16 cm in diameter is placed on a saline solution phantom - Fig. 7a - and then a lens of 3 cm in thickness is placed between the coil and the phantom - Fig. 7b. The MR images were obtained in a Siemens Avanto 1.5T system (Siemens Medical Systems, Erlangen, Germany) sited at the Department of Experimental Physics 5 (Biophysics) of the University of Würzburg (Würzburg, Germany). The coil was used in transmit/receive mode of operation. The pulse sequence used was a typical rf field mapping sequence, i.e., a high flip angle preparation pulse followed by a rapid image acquisition module.. This allows to visualize the field lines pattern produced by the coil. A field line in the MR image shown in Fig. 7a has been marked with a cross. The equivalent field line in Fig. 7b appears shifted into the phantom a distance of 3 cm, which is the thickness of the lens. This experiment clearly illustrates the ability of the metamaterial lens to translate the field profile of the coil deeper into the sample. Next, the general conclusions arising from the above analysis will be tested by means of experiments. The results shown in Fig. 4 are in agreement with the results of the previous experiment reported by the authors [9]. It is of interest to illustrate the present work with other MR images corresponding to a different coil geometry, and to include in the results a quantitative evaluation of the increase in the SNR, which was not provided in [9]. Fig. 8 shows a sketch of the experimental setup and the MR images obtained for this purpose. The MR images were obtained in a Siemens Simphony 1.5T system (Siemens Medical Systems, Erlangen, Germany) sited at Virgen Macarena University Hospital (Seville, Spain). One of the elements of a double loop array coil (Siemens Medical Systems, Erlangen, Germany) was used as detector. This array is usually applied for imaging of the temporo mandibular joints, eyes and wrists, and it was used in the experiments for imaging the ankles of one of the authors. A fat-supression pulse sequence with FOV= 219 x 250 mm$^2$, data matrix 224 × 256, TE=7.7 ms, TR=371 ms and slice thickness 3 mm was used and axial MR images were obtained. The comparison between the MR images shows that the presence of the lens increases the SNR in the ankle which is far from the coil. An estimation of the SNR was derived from the ratio between the mean signal to the standard deviation (SD) of a small circular ROI placed as shown in the figure. The ratio Mean/SD in the presence of the lens was 56.3/5.1 = 11 and it was 15.7/3.7 = 4.2 without the lens, that is, the SNR provided by the lens in the observed ROI was 2.6 times larger than without the lens. Therefore, there exists a qualitative agreement between these experimental results and the theoretical results shown in Fig. 4 (quantitative predictions of the SNR are out of the scope of this work since noise depends on the specific structure of the tissues, which can not be accounted for quantitatively by the simple model developed in the previous section). Next, an experiment was designed to validate the conclusion arising from the analysis shown in Fig. 5 for the combination of a lens with a magnetic wall. A theoretical magnetic wall is given by a semiinfinite medium with an infinite permeability. In practice, a magnetic wall can be implemented by means of a thick slab exhibiting a high permeability. In a first approach we simply fabricated a 2D array of split rings corresponding to the outer interface of this metamaterial slab. Since magnetic fields do not penetrate into the infinite permeability medium, it can be guess that this single layer will mimic to some extent the abovementioned metamaterial slab [11]. A photograph of this device, which has been termed "reflector" by the authors, is shown in Fig. 1 besides the lens. Fig. 9 shows a sketch of the experiment, as well as the axial and sagittal MR images obtained with the lens and the reflector. The pulse sequence was the same as that used for the results shown in the previous figure. With the lens and the reflector an increasing of the SNR is observed in the ankle placed between the coil and the lens. The ratio Mean/SD in the circular ROI indicated in the axial images corresponds to 30.9/4.6 = 6.7 for the case without lens and 62.4/7.7 = 8.1 for the case where the reflector-backed lens is used, so that this combined device provides a relative increasing of 20% in the SNR at that point. This result can be probably improved by optimizing the implementation of the magnetic wall (for example, by using a 3D array of split rings exhibiting a high permeability instead of a single 2D array). Anyway, the authors think that the reported experiment clearly shows the validity of the proposed concept. Finally, an experiment was carried out in order to check the compatibility of the lens with phased arrays. The lens used in the former experiments had two unit cells in depth and a thickness of 3 cm. The lens used for the present experiment had three unit cells in depth and a thickness of 4.5 cm. For this experiment, a bilateral four channel phased array coil (Machnet BV;

The Netherlands) was used in a Siemens Avanto 1.5T system (Siemens Medical Solutions, Erlangen, Germany) in the University Hospital Charleroi, Belgium. Axial images of the knees of a volunteer were obtained by using a T1-weighted spin-echo sequence with TR 450 ms, TE 12 ms and slices of 6 mm thickness, FOV of 189 × 319 mm2 and a 228x384 matrix. Fig. 10 shows a sketch of the experimental setup when the lens is present and when the lens is absent, and the T1-weighted images corresponding to both cases. When the lens was present, medial structures such as the vastus medialis and sartorius muscles in both legs are more clearly demonstrated and a quantitative increase in the SNR of 40% was measured. We feel that this experiment shows the potential usefulness of metamaterial lenses in order to improve the images of pairs of organs, such as knees, wrists or breasts.

## 4. Conclusions

Along this paper a theoretical model for the analysis of the sensitivity, the SNR and the FOV of MR coils in the presence of $\mu = -1$ metamaterial lenses has been developed. Our analysis has shown that metamaterial lenses usually improve the signal and the localization of the FOV of surface coils. However, the SNR is only improved if the lens is placed at a distance from the coil which, as a rule of thumb, should be larger than the coil diameter. A combination of a lens with a metamaterial mimicking a magnetic wall may behave as a reflector, increasing the signal received by a coil located at the opposite side of a given organ or tissue. These conclusions have been checked successfully by "in situ" MR experiments. Although the purpose of these experiments was only to prove the concepts developed along the text, the results suggest some applications of clinical interest that would require further investigation and that would be oriented, for example, to image pairs of joints simultaneously. Although the SNR could be increased in a conventional way by introducing more active coils between the joints, the lens allows to get this increasing by means of a passive device which does not require addition of active channels. These results encourage us to investigate the possibility of using the lens to image the female breast in a similar way as it has been done with joints. Higher field strengths have been used to improve breast diagnostic capabilities [20, 21]. However, due to the risk of generating hotspots during MR imaging with higher fields, standard absorption rate (SAR) limits are being enforced by the U.S. Food and Drug Administration and other healthcare safety groups, and only the development of more advanced coil concepts could allow to address the issue [21]. Metamaterials could provide a different way to increase the SNR by fulfilling the safety regulations governing SAR. A second application for metamaterial lenses is related with the techniques of parallel imaging. Thus, for example, in the PILS technique [19] it is essential that the FOV of each coil is well localized and restricted to a finite region of space. However, in conventional coil arrays, this localization takes place only at distances close to the array because the field produced by the coils spread out at farther distances. Since the lens can help to discriminate the fields produced by individual coils without a substantial loss in the SNR, the lens could be used to increase the penetration depth of the array of coils in the sense that it would be possible to get localized field patterns for each coil at longer distances from the array. Computer simulations carried out by means of the method described in this paper support this conclusion. In summary, the authors feel that, in general, the emerging technology of metamaterials could help to improve several aspects of MR imaging by providing new concepts for the advancement of the MR technology, and that split ring $\mu = -1$ metamaterial lenses could play an important role in this direction.

## 5. Acknowledgments

This work has been supported by the Spanish Ministerio de Educación y Ciencia under Project No. TEC2007-68013-C02-01/TCMand by the Spanish Junta de Andalucía under Project No. P06-TIC-01368. Lukas Jelinek also thanks for the support of the Czech Grant Agency (project no. 102/08/0314). We want to thank Dr. Carlos Caparr´os, from Virgen Macarena University Hospital (Seville, Spain), Dr. Bavo van Riet, MR Regional Business Manager for South-West Europe from SIEMENS Medical Solutions and his partnersRoger Demeure and Pierre Foucart from Charleroi University Hospital (Charleroi, Belgium), and Prof. Peter M. Jakob and Dr. Volker C. Behr, from the Department of Experimental Physics 5 (Biophysics) of the University of Würzburg (Germany) for providing the MRI facilities used in this work, and also for helpful discussions. We want also to thank Dr. Francisco Moya, from PET Cartuja (Seville, Spain), for his advice and support.

## 6. Figure legends

**Figure 1.** Photograph of the metamaterial lens and reflector. The lens consists of a 3D array of capacitively-loaded copper rings with $18 \times 18 \times 2$ cubic unit cells containing a total of 2196 rings. It has been designed to exhibit a permeability equal to -1 at the Larmor frequency of 63.87 MHz in a 1.5T system. The reflector consists of a 2D array of $14 \times 14$ rings which constitutes the first layer of a high permeability metamaterial slab.

**Figure 2.** Model for the analysis of the sensitivity and the SNR of a circular coil in presence of a metamaterial slab and one or more samples. The coil is modelled by a flat perfect conducting strip. The complex permittivity and/or permeability of each layer of thickness t models either the metamaterial lens or the tissue.

**Figure 3.** On the left side it is shown a sketch of the configurations which are theoretically analyzed, a coil of 3 inch in diameter with a strip width of 1 cm at certain distance d of a semi-infinite lossy media and the coil at the same distance from a lens, which is placed in contact with the lossy media. On the right side it is shown the calculation of the magnetic field per unit current along the coil axis in these configurations for (a) d = 1 cm and (b) d = 6 cm. The SNR of the coil is also shown for (c) d = 1 cm and (d) d = 6 cm. The lens is modelled as a slab with a complex permeability $\mu = -1 - j0.25$ and the lossy media has a complex permittivity $\varepsilon = 90 - j197$. The frequency is 63.87 MHz.

**Figure 4.** (a) SNR along the axis of a coil placed in front of two lossy slabs of 10 cm of thickness separated by a lens of 3 cm. A 2D plot of the SNR in a plane perpendicular to the coil and containing its axis is shown for the slab placed beyond the lens when (b) the lens is present and when (c) the lens has been replaced by an air layer of 3 cm.

**Figure 5.** (a) SNR along the axis of a coil with a lossy slab of 10 cm of thickness placed between the coil and the lens backed by a reflector. A 2D plot of the SNR in a plane perpendicular to the coil and containing its axis is shown for the slab when (b) the lens backed by the reflector is present and when (c) the lens and the reflector are absent.

**Figure 6.** Plots of the calculated sensitivity at 6 cm inside a lossy semiinfinite slab (" = 90−j197) for two rectangular coils with dimensions of 7×23 cm - dimensions of the coil geometry used for PILS demonstration [19] – and with their centers separated 10 cm. (a) Sketch of the configuration. (b) and (c ) images when the coils are placed at 1.5 cm from (a) the sample and (b) when they are placed at the same distance from the lens. The field pattern of each coil can be distinguished much better when the lens is present, which suggests the use of the lens for parallel imaging.

**Figure 7**. MR images obtained for a circular coil of 16 cm in diameter by using a RF field sequence when (a) it is placed on a water saline phantom and (b) when a 3 cm thick lens is placed between the coil and the phantom. The field lines in case (a) appear in case (b) shifted inside the phantom a distance which is similar to the thickness of the lens.

**Figure 8.** Sketch of the experiment carried out in a 1.5T Simphony system from SIEMENS (Siemens Medical Solutions, Erlangen, Germany) in the Virgen Macarena University Hospital (Seville, Spain) using an element of a double loop array coil (Siemens Medical Systems, Erlangen, Germany) as detector. A fat-supression pulse sequence with FOV=25 cm, data matrix 224 × 256, TE=7.7 ms, TR=371 ms, and slice thickness of 3 mm was used. Axial MR images of the ankles of one of the authors were obtained with and without the lens. A small ROI for determination and comparison of the SNR in both situations is shown in the images. The SNR was increased 2.6 times with the lens.

**Figure 9**. With the same experimental setup as in Fig. 5, axial and sagittal MR images were obtained by using the lens backed by the reflector. The images show that the SNR is increased in the ankle placed between the detector and the reflector-backed lens. Direct measurements of the SNR in the small ROI indicated in the images show an increase of 20% by using the lens and the reflector.

**Figure 10.** Sketch of the experiment carried out in a 1.5T Avanto system from SIEMENS (Siemens Medical Solutions, Erlangen, Germany) in the University Hospital Charleroi (Charleroi, Belgium) using a bilateral four channel phased array coil (Machnet BV; The Netherlands) and MR images obtained (axial T1-weighted spin-echo sequences with TR 450 ms, TE 12 ms). The knees are placed between the elements of the array and the distance between the knees is equal to the thickness of the lens (4.5 cm). (a) As expected the signal is stronger near the surface coil and more superficial structures are better represented than deeper ones due to the longer distance to the coil.
(b) The lens is placed between the knees and then medial structures such as the vastus medialis and sartorius muscles in both legs are more clearly shown.

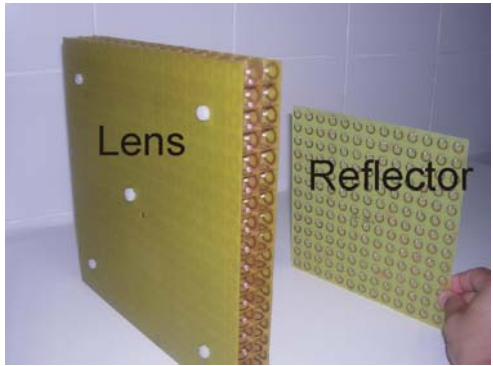

**Figure 1**

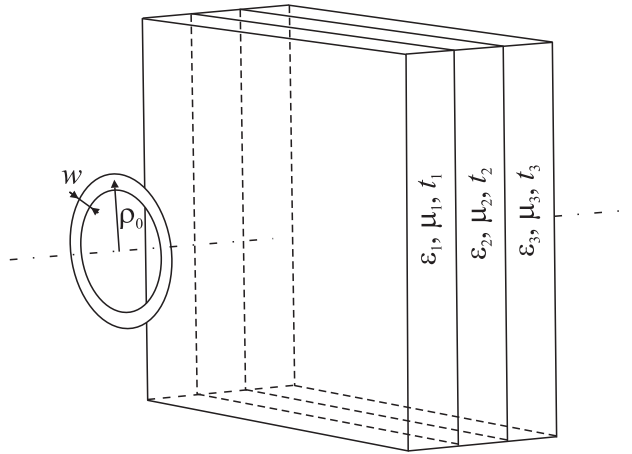

**Figure 2**

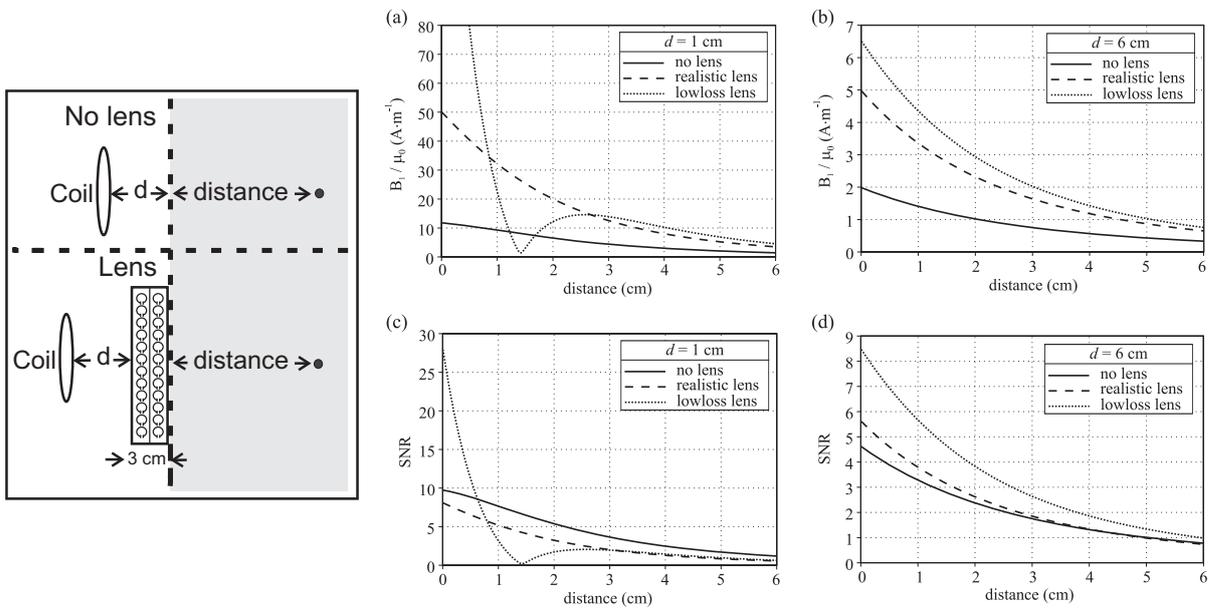

**Figure 3**

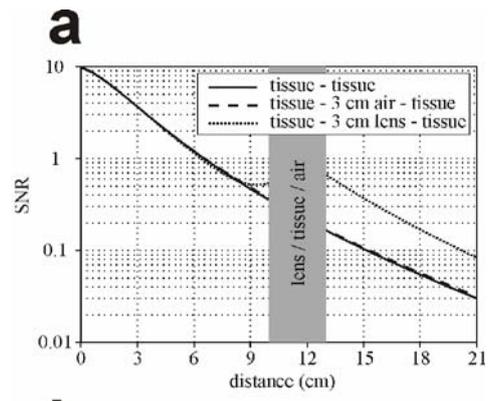
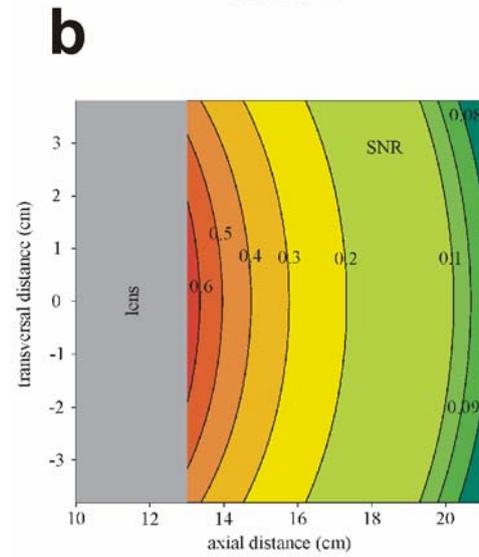
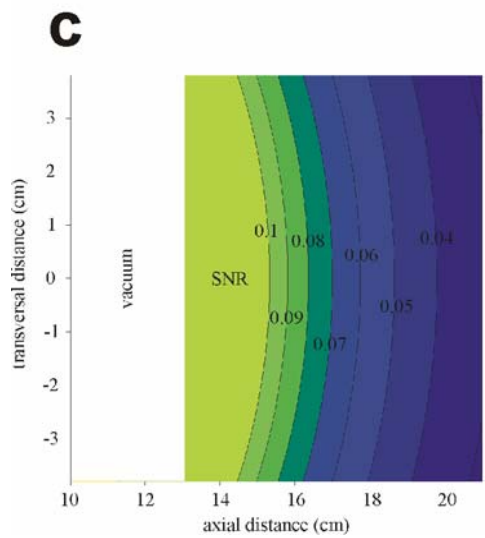

**Figure 4**

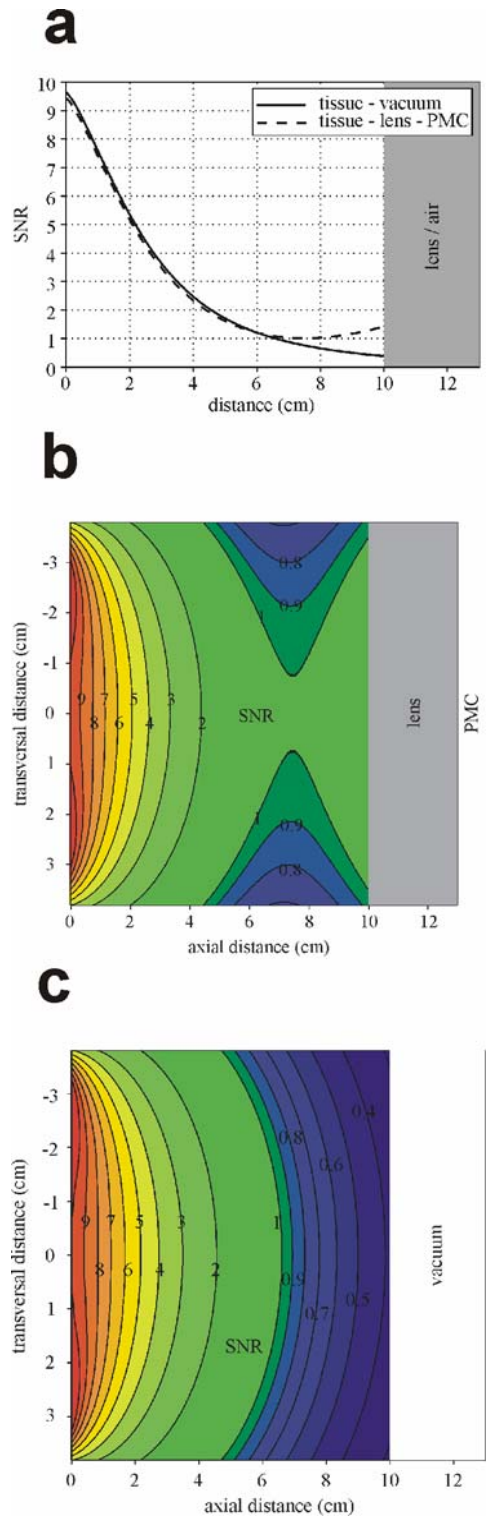

**Figure 5**

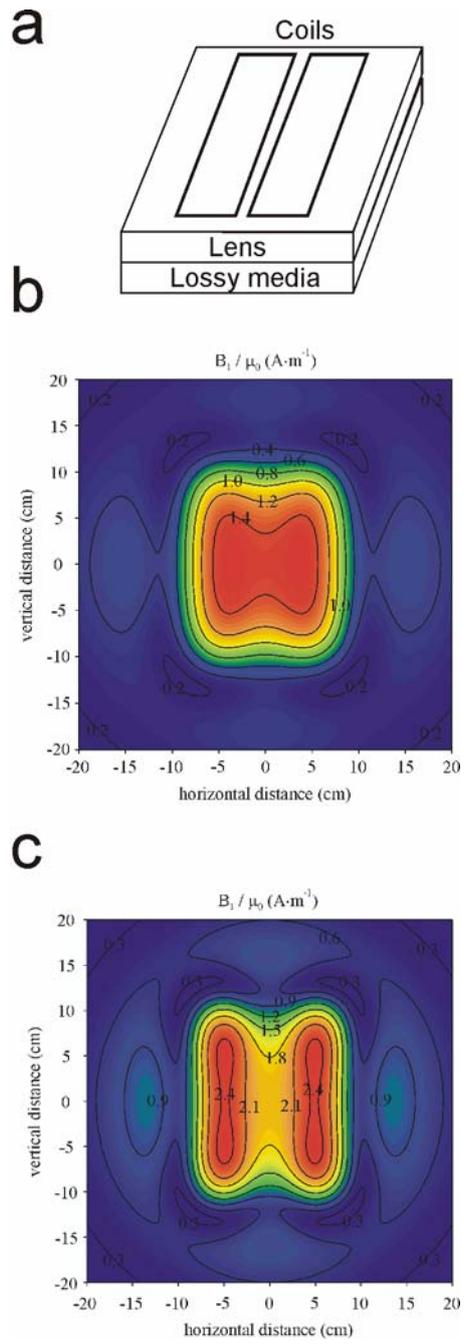

**Figure 6**

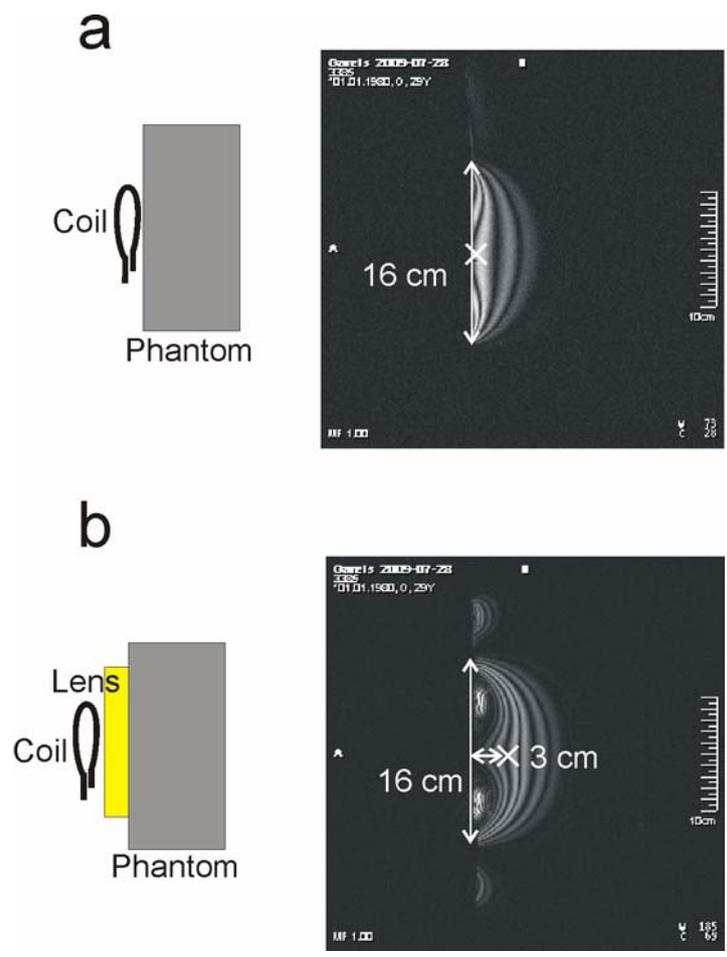

**Figure 7**

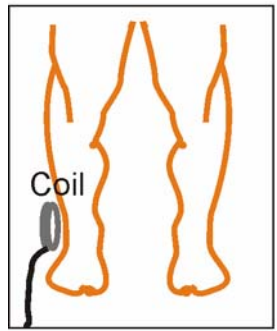
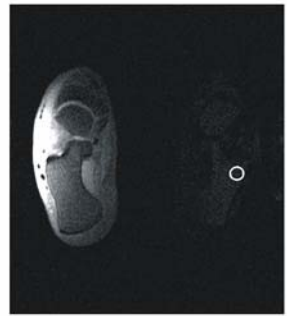
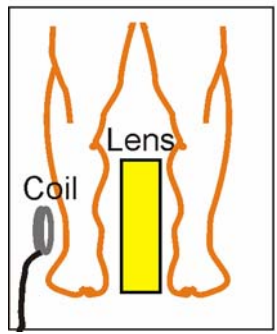
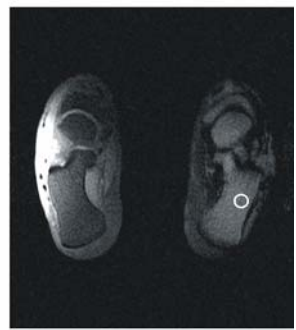

**Figure 8**

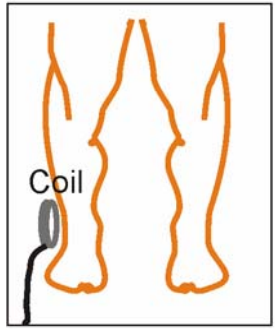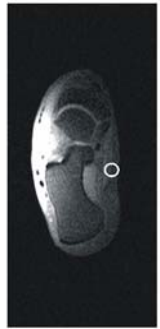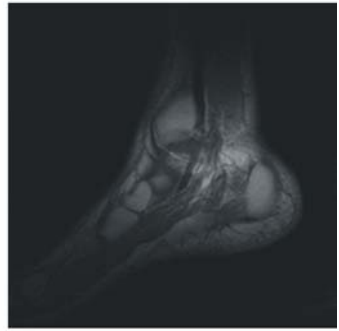
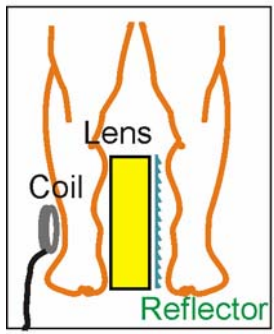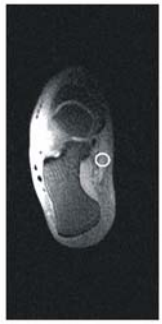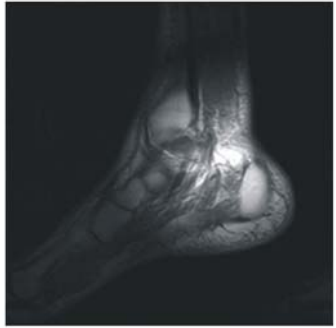

**Figure 9**

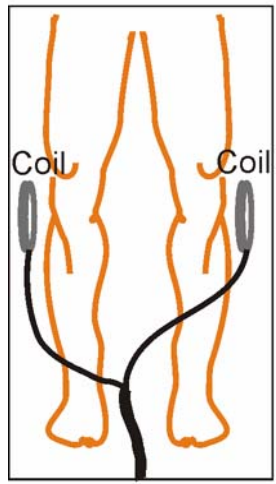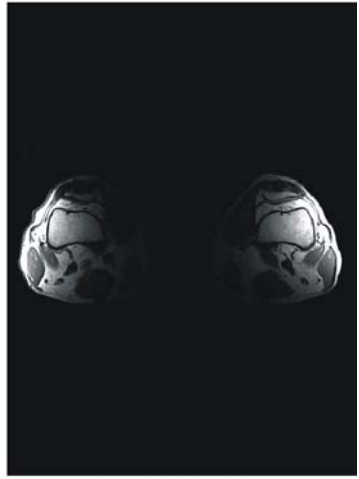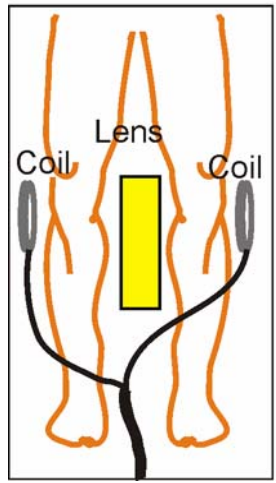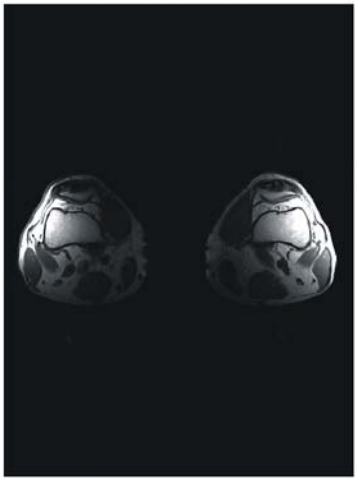

**Figure 10**